\documentclass[11pt]{article}
\usepackage{amsmath,amssymb,epsf}

\newcommand{\ZZ}{\mathbb{Z}}
\newcommand{\QQ}{\mathbb{Q}\hspace{0.5pt}}
\newcommand{\RR}{\mathbb{R}}
\newcommand{\CC}{\mathbb{C}}

\begin{document}

\begin{center}
\vspace*{0.2in}\begin{LARGE}\begin{bf}
Averaged coordination numbers of\\[1mm] planar aperiodic 
tilings
\end{bf}
\end{LARGE}
\vspace{8mm}

{\large MICHAEL~BAAKE$^{\dag}$ and
UWE~GRIMM$^{\ddag}$\footnote{Corresponding author. 
Email: u.g.grimm@open.ac.uk}}\vspace{3mm} 

$\mbox{}^{\dag}$Fakult\"{a}t f\"{u}r Mathematik, 
Universit\"{a}t Bielefeld,\\ Postfach 100131, 33501 Bielefeld, 
Germany\\    
$\mbox{}^{\ddag}$Applied Mathematics Department, 
The Open University,\\ Walton Hall, Milton Keynes MK7 6AA, UK
\end{center}\vspace{8mm}

\noindent\hspace{0.4in}\begin{small}\parbox{5in}{ 
We consider averaged shelling and coordination numbers of aperiodic
tilings. Shelling numbers count the vertices on radial shells around a
vertex. Coordination numbers, in turn, count the vertices on
coordination shells of a vertex, defined via the graph distance given
by the tiling.  For the Ammann-Beenker tiling, we find that
coordination shells consist of complete shelling orbits, which enables
us to calculate averaged coordination numbers for rather large
distances explicitly. The relation to topological invariants of
tilings is briefly discussed.
\vspace{8mm}

}\end{small}
\vspace{1cm}

\subsection{Introduction}

Many combinatorial questions from lattice theory are best extended to
aperiodic system by using an additional averaging process. In
particular, this is the case for the shelling problem, where one asks
for the number of vertices on spherical (circular) shells. In this
contribution, we consider an extension of the shelling problem to the
setting of more general distances, and give examples for the
coordination number case \cite{BG1,BGRJ}, which corresponds to the
graph distance in a tiling.

To keep things simple, we explain our approach for cyclotomic model
sets in the Euclidean plane $\RR^2\simeq\CC$, with co-dimension $2$
(referring to the so-called internal space). Here, following the
algebraic setting of Pleasants \cite{P}, one starts from a set of
cyclotomic integers, $L=\ZZ[\xi_{n}]$ with $\xi_{n}=e^{2\pi i/n}$ and
suitable $n$, which is the set of all integer linear combinations of
the regular $n$-star of unit length. Note that one can choose a
$\ZZ$-basis of $\phi(n)$ elements, where $\phi$ is Euler's totient
function \cite{A}.

\begin{figure}
\centerline{\epsfxsize=0.4\textwidth\epsfbox{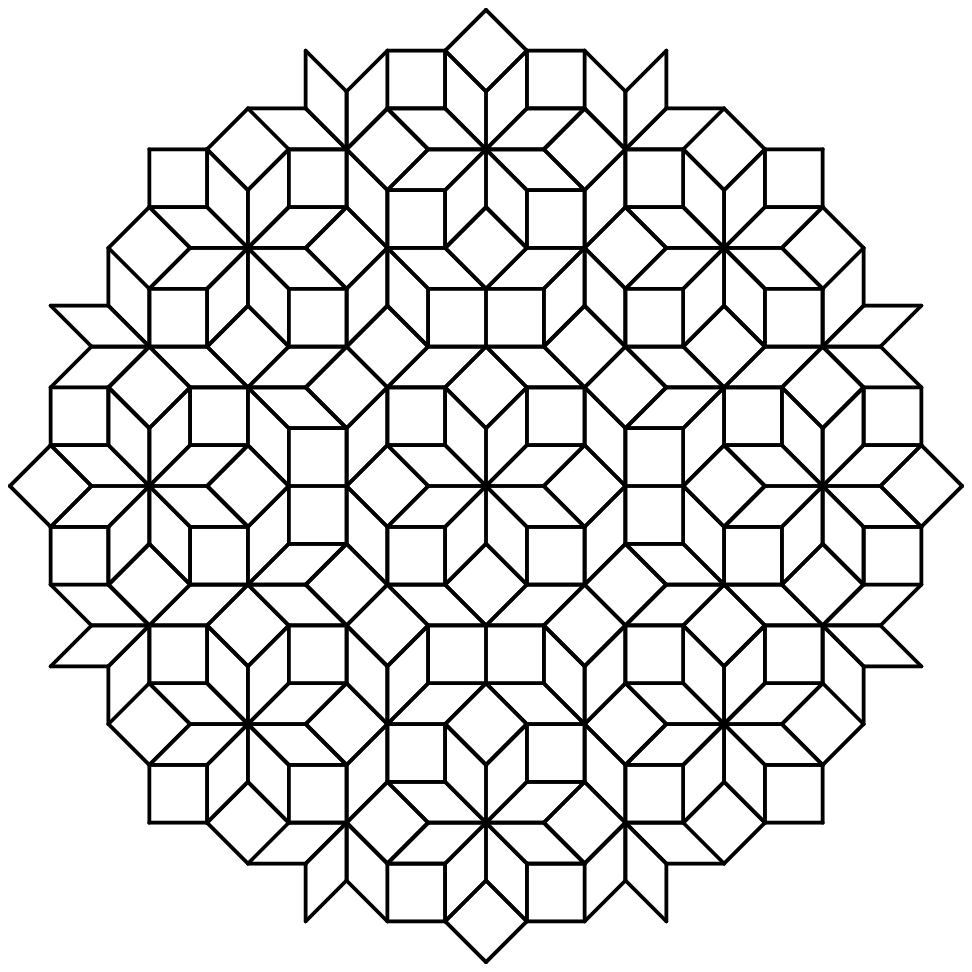}\hspace{0.075\textwidth}
        \epsfxsize=0.4\textwidth\epsfbox{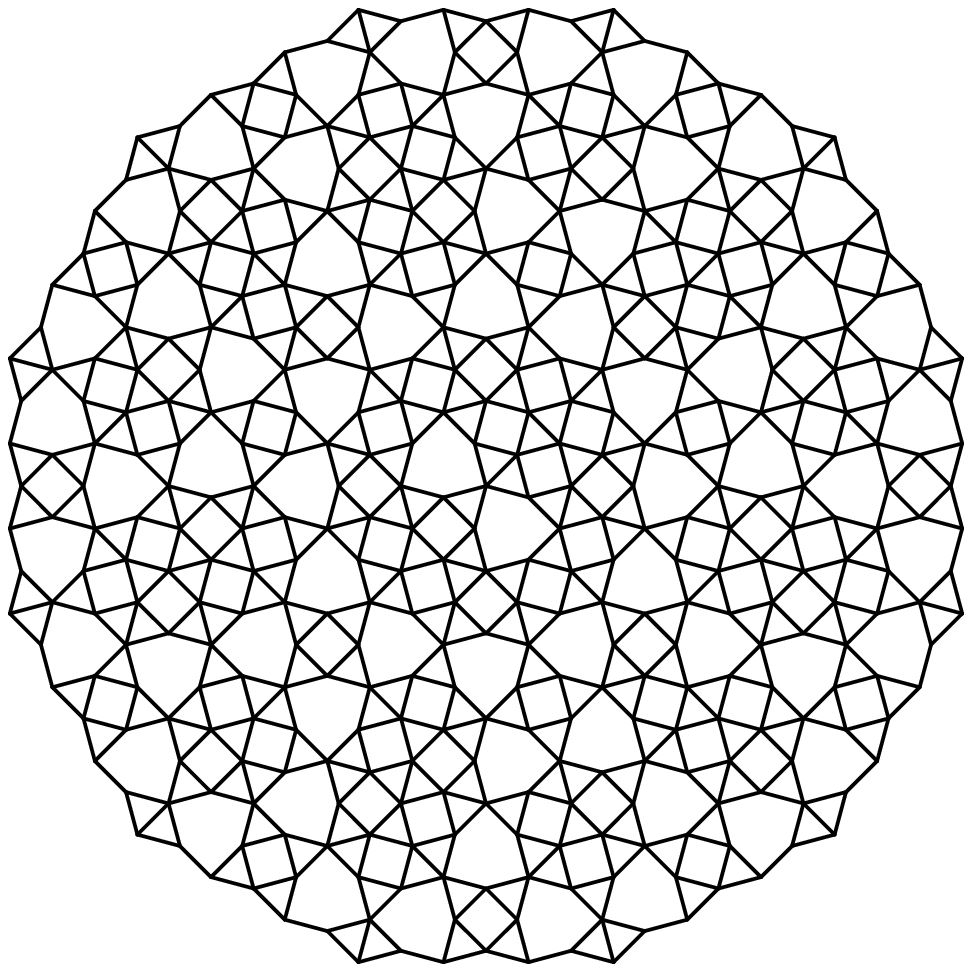}}
\caption{Patches of the Ammann-Beenker tiling (left) and 
the shield tiling (right).\label{fig:til}}
\end{figure}

This setting is equipped with a natural $\star$-map, defined by a
suitable algebraic conjugation (such as
$\xi_{5}^{}\mapsto\xi_{5}^{2}$, $\xi_{8}^{}\mapsto\xi_{8}^{3}$, and
$\xi_{12}^{}\mapsto\xi_{12}^{5}$ in the examples discussed below,
together with the canonical extension to all elements of
$L$). The set $\tilde{L} := \{ (x,x^{\star}) \mid x\in L\}$
is then a {\em lattice}\/ in $\RR^{\phi(n)}$, the so-called Minkowski
embedding \cite{BS} of $L$. A \emph{model set}\/
$\varLambda$ is now a set of the form
\begin{equation}
\varLambda = \{x\in L\mid x^{\star}\in\varOmega\},
\end{equation}
or any translate of it, where the window $\varOmega$ is a relatively
compact subset of internal space with non-empty interior. A natural
choice for $\varOmega$ that preserves $n$-fold symmetry is a regular
$n$-gon, which leads to \emph{regular}\/ model sets (i.e., the
boundary $\partial\varOmega$ has Lebesgue measure $0$). More
precisely, we focus on the \emph{generic}\/ case
($\partial\varOmega\cap L^{\star}=\varnothing$), where $\varLambda$ is
repetitive and its LI-class (consisting of all locally
indistinguishable patterns) defines a uniquely ergodic dynamical
system \cite{Martin}. This includes the examples of figure~\ref{fig:til}.

Let $d(x,y)$ denote any translation invariant distance between $x$ and
$y$.  Due to unique ergodicity, combined with finite local complexity,
we know \cite{Martin,BG3} that the averaged number $s^{}_{d}(r)$ of
vertices on a $d$-shell of radius $r$ is determined by a sum over a
{\em finite}\/ number of patches, weighted by their frequencies, which
exist uniformly. Since we work with model sets, this can be further
reduced to a sum that involves only admissible pairs of vertices with
the correct distance. The frequency of a pair $(x,y)$ with $x,y\in L$
is given by the scale-independent autocorrelation coefficient
$\nu(x-y)$, where \cite{BG3}
\begin{equation}
\nu(z)\;=\;\frac{1}{\mathrm{vol}(\varOmega)}
\int_{\mathbb{R}^{m}} 
\mathbf{1}^{}_{\varOmega}(w)\,
\mathbf{1}^{}_{\varOmega}(w+z^{\star})\,\mathrm{d}w,
\label{eq:auto}
\end{equation}
with $\mathbf{1}^{}_{\varOmega}$ denoting the characteristic function
of the window.  The normalisation is such that $\nu(0)=1$, i.e., we
count per point of $\varLambda$ rather than per unit volume. Clearly,
$\nu(z)=\nu(-z)$, and further identities may occur as a result of the
symmetries of the window. In general, the averaged number reads
\begin{equation}
s^{}_{d}(r)\; = \sum_{\substack{z\in\varLambda-\varLambda \\ d(0,z)=r}} 
\nu(z),
\label{eq:shell}
\end{equation}
which can then be further simplified by means of a standard orbit
analysis \cite{BG3}. 

In view of this derivation, it is reasonable to consider a $d$-shell
to be the collection of all pairs $(x,y)$ of a given distance 
$d(x,y)$ {\em together}\/ with their frequencies $\nu(x-y)$.

The coordination problem is now the extension of the shelling problem
to a different distance concept, based on the graph distance in the
tiling under consideration. The graph distance of two vertices is
defined as the minimum number of edges in a path linking the two
vertices. For simplicity, we restrict our discussion to examples with
a single edge type (i.e., all edges have the same length), though
various extensions are possible.

There is an important connection between averaged coordination and
shelling numbers, which stems from the relation between graph and
Euclidean distances. On a given coordination shell, the vertices
appear in symmetry orbits of the underlying tiling; these vertices are
orbitwise distributed over finitely many Euclidean shells. Within one
orbit, every vertex contributes the same amount to the averaged
numbers.  Conversely, the vertices on a Euclidean shell orbitwise
belong to (possibly different) coordination shells. Here,
the term orbit simply refers to the orbit of a point under the point
symmetry group of the pattern under consideration.

\subsection{Examples}

In what follows, we concentrate on three particular examples, with 8-,
10-, and 12-fold symmetry. For $n=8$, we consider the Ammann-Beenker
tiling, obtained by the above construction with a regular octagon of
edge length $1$ as the window. The special role of $\sqrt{2}$ reflects
the fact that $\ZZ[\sqrt{2}]=\RR\cap\ZZ[\xi_{8}]$, see \cite{BG3} for
details. This reference also contains the known results on the
circular shelling.

\begin{table}
\begin{small}
\caption{Averaged coordination numbers of the 
Ammann-Beenker tiling.\label{tab:ab}}
\begin{center}
\begin{tabular}{rr@{$\;$}c@{$\;$}rr}
\hline
\multicolumn{1}{c}{$k$} & 
\multicolumn{3}{c}{$s_{\mathrm c}(k)$} & 
\multicolumn{1}{c}{num.~value}\rule[-1.5ex]{0ex}{4.5ex}\\
\hline
 $1$ & $4$&&                        &   $4.000$\rule[0ex]{0ex}{2.5ex} \\
 $2$ & $32$&$-$&$16\sqrt{2}$        &   $9.373$ \\
 $3$ & $-8$&$+$&$16\sqrt{2}$        &  $14.627$ \\
 $4$ & $24$&$-$&$4\sqrt{2}$         &  $18.343$ \\
 $5$ & $40$&$-$&$12\sqrt{2}$        &  $23.029$ \\
 $6$ & $40$&$-$&$8\sqrt{2}$         &  $28.686$ \\
 $7$ & $-176$&$+$&$148\sqrt{2}$     &  $33.304$ \\
 $8$ & $444$&$-$&$288\sqrt{2}$      &  $36.706$ \\
 $9$ & $240$&$-$&$140\sqrt{2}$      &  $42.010$ \\
$10$ & $-648$&$+$&$492\sqrt{2}$     &  $47.793$ \\[1ex]
$11$ & $232$&$-$&$128\sqrt{2}$      &  $50.981$ \\
$12$ & $508$&$-$&$320\sqrt{2}$      &  $55.452$ \\
$13$ & $-272$&$+$&$236\sqrt{2}$     &  $61.754$ \\
$14$ & $-556$&$+$&$440\sqrt{2}$     &  $66.254$ \\
$15$ & $1540$&$-$&$1040\sqrt{2}$    &  $69.218$ \\
$16$ & $980$&$-$&$640\sqrt{2}$      &  $74.903$ \\
$17$ & $-3064$&$+$&$2224\sqrt{2}$   &  $81.211$ \\
$18$ & $1424$&$-$&$948\sqrt{2}$     &  $83.326$ \\
$19$ & $812$&$-$&$512\sqrt{2}$      &  $87.923$ \\
$20$ & $740$&$-$&$456\sqrt{2}$      &  $95.119$\rule[-1ex]{0ex}{1ex}\\
\hline
\end{tabular}\hspace{2ex}
\begin{tabular}{rr@{$\;$}c@{$\;$}rr}
\hline
\multicolumn{1}{c}{$k$} & 
\multicolumn{3}{c}{$s_{\mathrm c}(k)$} & 
\multicolumn{1}{c}{num.~value}\rule[-1.5ex]{0ex}{4.5ex}\\
\hline
$21$ & $-3284$&$+$&$2392\sqrt{2}$   &  $98.799$\rule[0ex]{0ex}{2.5ex} \\
$22$ & $2172$&$-$&$1464\sqrt{2}$    & $101.591$ \\
$23$ & $4164$&$-$&$2868\sqrt{2}$    & $108.036$ \\
$24$ & $-8648$&$+$&$6196\sqrt{2}$   & $114.467$ \\
$25$ & $6836$&$-$&$4752\sqrt{2}$    & $115.657$ \\
$26$ & $3164$&$-$&$2152\sqrt{2}$    & $120.612$ \\
$27$ & $-7972$&$+$&$5728\sqrt{2}$   & $128.615$ \\
$28$ & $1500$&$-$&$968\sqrt{2}$     & $131.041$ \\
$29$ & $4716$&$-$&$3240\sqrt{2}$    & $133.948$ \\
$30$ & $792$&$-$&$460\sqrt{2}$      & $141.462$ \\[1ex]
$31$ & $-10216$&$+$&$7328\sqrt{2}$  & $147.357$ \\
$32$ & $10500$&$-$&$7320\sqrt{2}$   & $147.957$\\
$33$ & $7236$&$-$&$5008\sqrt{2}$    & $153.618$\\
$34$ & $-18132$&$+$&$12936\sqrt{2}$ & $162.267$\\
$35$ & $5356$&$-$&$3672\sqrt{2}$    & $163.008$\\
$36$ & $7328$&$-$&$5064\sqrt{2}$    & $166.423$\\
$37$ & $2800$&$-$&$1856\sqrt{2}$    & $175.220$\\
$38$ & $-19444$&$+$&$13876\sqrt{2}$ & $179.627$\\
$39$ & $12416$&$-$&$8652\sqrt{2}$   & $180.224$\\
$40$ & $21932$&$-$&$15376\sqrt{2}$  & $187.052$\rule[-1ex]{0ex}{1ex}\\
\hline
\end{tabular}
\end{center}
\end{small}
\end{table}

For this tiling, the connection between shelling and coordination
numbers is rather advantageous, because coordination shells comprise
only \emph{complete}\/ circular shells. This is a consequence of the
fact that the four directions of the edges in the tiling form a
$\ZZ$-basis of the underlying module $\ZZ[\xi_{8}]$, which is possible
because $\phi(n)=n/2$ for $n=8$. This means that for a given distance
vector, the number of steps along each direction is uniquely
determined. Moreover, there always exists at least one path along
edges of the actual tiling that is admissible (in the sense that never
has to `backtrack' along the path).  This is a higher-dimensional
analogue of the corresponding (trivial) situation for the silver mean
chain (when viewed as a cut and project set obtained from a
rectangular lattice), which fails for other tilings.

\begin{figure}
\centerline{\epsfxsize=0.45\textwidth\epsfbox{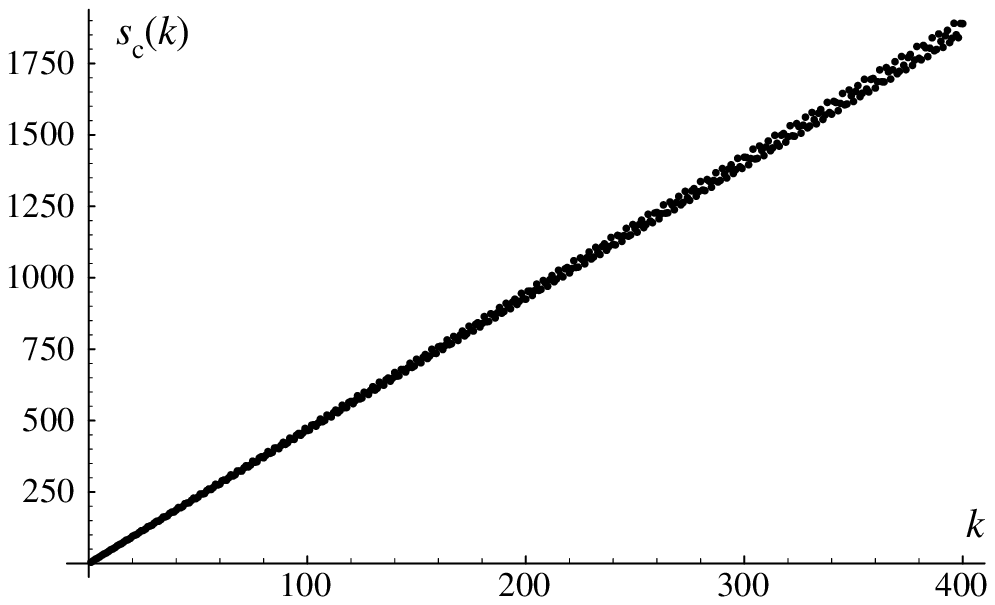}
   \hspace{0.05\textwidth}\epsfxsize=0.45\textwidth\epsfbox{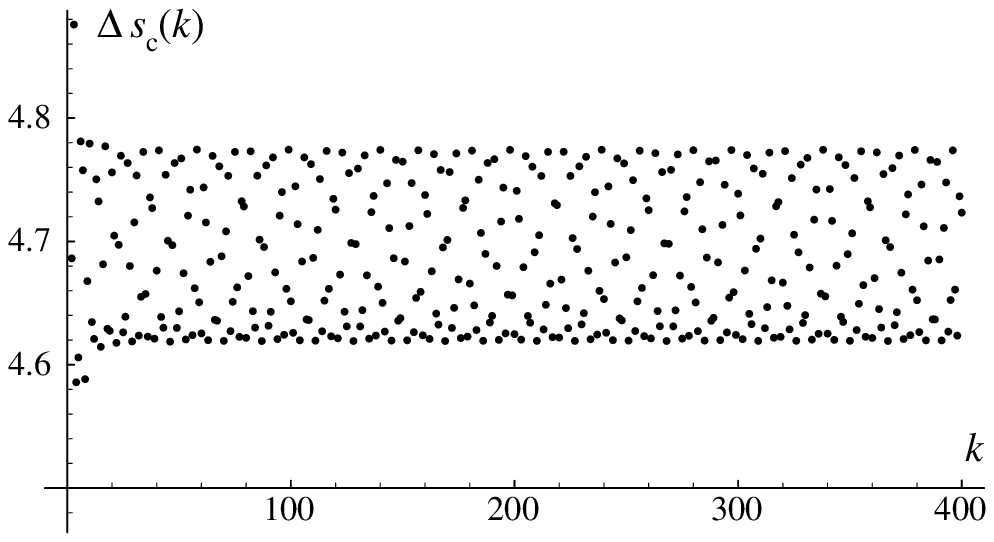}}
\caption{The first 400 coordination numbers for the Ammann-Beenker 
tiling (left) and the fluctuations in the differences of successive 
coordination numbers (right).\label{fig:ab}}
\end{figure}

The averaged coordination numbers can therefore be calculated by
identifying the contributing circular shells, and summing the
corresponding averaged shelling numbers, which can be obtained as
described in \cite{BG3}. Following this approach, we calculated the
first few hundred averaged coordination numbers for the Ammann-Beenker
tiling, see table~\ref{tab:ab} and figure~\ref{fig:ab}.

As for periodic planar lattices, the averaged coordination numbers
$s_{\mathrm{c}}(k)$ grow, on the average, linearly with the number of
steps $k$. However, a closer inspection of the left part of
figure~\ref{fig:ab} shows that the growth rate fluctuates, and that
the data points do not lie on a single line, but inside a sector
bounded by two lines of slightly different slopes. An even closer
inspection reveals that the fluctuations of $s_{\mathrm{c}}(k)$ follow
a sophisticated pattern, as displayed in the right part of
figure~\ref{fig:ab} which shows the differences $\Delta
s_{\mathrm{c}}(k)= s_{\mathrm{c}}(k+1)-s_{\mathrm{c}}(k)$ of
consecutive averaged shelling numbers. The resulting pattern appears
to comprise a number of sinusoidally varying curves. This might be
caused by the variations of overlap areas of the window with shifted
copies of itself, which enters the computation of patch frequencies; a
closer investigation of this phenomenon might lead to interesting
results.

\smallskip

A construction of the rhombic Penrose tiling as a cyclotomic model set
with four components is described in \cite{BM}. Due to the necessity
of four windows (which are also present in the non-minimal embedding
via the lattice $\ZZ^5$), the determination of averaged quantities is
technically more involved. Table~\ref{tab:pen} recalls some results
obtained earlier in \cite{BGRJ}. In this case,
$\ZZ[\xi_{5}]\cap\RR=\ZZ[\tau]$, where $\tau=(1+\sqrt{5})/2$ is the
golden ratio, which is the relevant algebraic integer here.

\begin{table}[b]
\begin{small}
\caption{Averaged coordination numbers of the rhombic 
Penrose tiling.\label{tab:pen}}
\begin{center}
\begin{tabular}{rr@{$\;$}c@{$\;$}rr}
\hline
\multicolumn{1}{c}{$k$} & 
\multicolumn{3}{c}{$s_{\mathrm c}(k)$} & 
\multicolumn{1}{c}{num.~value}\rule[-1.5ex]{0ex}{4.5ex}\\ 
\hline
 $1$ & $4$&&                  &  $4.000$\rule[0ex]{0ex}{2.5ex} \\
 $2$ & $58$&$-$&$30\tau$      &  $9.459$ \\
 $3$ & $-128$&$+$&$88\tau$    & $14.387$ \\
 $4$ & $288$&$-$&$166\tau$    & $19.406$ \\
 $5$ & $-374$&$+$&$246\tau$   & $24.036$\rule[-1ex]{0ex}{1ex}\\
\hline
\end{tabular}\qquad
\begin{tabular}{rr@{$\;$}c@{$\;$}rr}
\hline
\multicolumn{1}{c}{$k$} & 
\multicolumn{3}{c}{$s_{\mathrm c}(k)$} & 
\multicolumn{1}{c}{num.~value}\rule[-1.5ex]{0ex}{4.5ex}\\ 
\hline
 $6$ & $980$&$-$&$588\tau$    & $28.596$\rule[0ex]{0ex}{2.5ex} \\
 $7$ & $-1614$&$+$&$1018\tau$ & $33.159$ \\
 $8$ & $2688$&$-$&$1638\tau$  & $37.660$ \\
 $9$ & $-3840$&$+$&$2400\tau$ & $43.282$ \\
$10$ & $4246$&$-$&$2594\tau$  & $48.820$\rule[-1ex]{0ex}{1ex}\\
\hline
\end{tabular}
\end{center}
\end{small}
\end{table}

\smallskip

As an example with 12-fold symmetry, we consider the so-called shield
tiling \cite{G} of figure~\ref{fig:til}. It is obtained from
$\ZZ[\xi_{12}]$ choosing a regular dodecagon of edge length $1$ as
window, see \cite{BG2} for details. In this case,
$\ZZ[\xi_{12}]\cap\RR=\ZZ[\sqrt{3}]$; examples are listed in
table~\ref{tab:shield}. Note that a single circular shell can
contribute to several coordination shells.

\begin{table}
\begin{small}
\caption{Averaged coordination numbers of the twelvefold symmetric
shield tiling, split into the contributions from circular shells of 
radius $r$.\label{tab:shield}}
\begin{center}
\begin{tabular}{rr@{$\;$}c@{$\;$}rrr@{$\;$}c@{$\;$}rr@{$\;$}c@{$\;$}r}
\hline
\multicolumn{1}{c}{$k$} & 
\multicolumn{3}{c}{$s_{\mathrm c}(k)$} & 
\multicolumn{1}{c}{\makebox[0pt]{num.~value}} &
\multicolumn{3}{c}{$r^2$} &
\multicolumn{3}{c}{contribution\rule[-1.5ex]{0ex}{4.5ex}}\\ 
\hline
 $1$ &  $8$&$-$&$2\sqrt{3}$ & $4.536$ & $2$&$-$&$\sqrt{3}$  
     & $8$&$-$&$2\sqrt{3}$ \rule[-1ex]{0ex}{3.5ex}\\
\hline
 $2$ & $20$&$-$&$6\sqrt{3}$ & $9.608$ & $4$&$-$&$2\sqrt{3}$
     & $2$&&\rule[-1.5ex]{0ex}{4ex}\\
     &     &   &            &              & $6$&$-$&$3\sqrt{3}$
     & $4$&$-$&$2\sqrt{3}$\\
     &     &   &            &              & $1$&&         
     & $14$&$-$&$4\sqrt{3}$\rule[-1ex]{0ex}{1ex}\\
\hline
 $3$ & $64$&$-$&$28\sqrt{3}$& $15.503$ & $1$&&          
     & $-6$&$+$&$4\sqrt{3}$\rule[0ex]{0ex}{2.5ex}\\
     &     &  &             &               & $5$&$-$&$2\sqrt{3}$
     & $10$&$-$&$4\sqrt{3}$\\
     &     &  &             &               & $2$&&          
     & $48$&$-$&$24\sqrt{3}$\\
     &     &  &             &               & $4$&$-$&$\sqrt{3}$
     & $12$&$-$&$4\sqrt{3}$\rule[-1ex]{0ex}{1ex}\\
\hline
 $4$ & $-46$&$+$&$38\sqrt{3}$ & $19.818$ & $4$&$-$&$\sqrt{3}$
     & $-6$&$+$&$4\sqrt{3}$\rule[0ex]{0ex}{2.5ex}\\
     &   &  &             &               & $8$&$-$&$3\sqrt{3}$
     & $-76$&$+$&$44\sqrt{3}$\\
     &   &  &             &               & $3$&&          
     & $-4$&$+$&$\frac{16}{3}\sqrt{3}$\\
     &   &  &             &               & $7$&$-$&$2\sqrt{3}$
     & $20$&$-$&$\frac{32}{3}\sqrt{3}$\\
     &   &  &             &               & $2$&$+$&$\sqrt{3}$ 
     & $12$&$-$&$2\sqrt{3}$\\
     &   &  &             &               & $4$&&          
     & $8$&$-$&$\frac{8}{3}\sqrt{3}$\rule[-1ex]{0ex}{1ex}\\
\hline
\end{tabular}
\end{center}
\end{small}
\end{table}

\subsection{Frequency modules}

It is remarkable that the averaged coordination numbers are special
algebraic integers in all three examples. From the cut and project
method, in conjunction with equations \eqref{eq:auto} and
\eqref{eq:shell}, it is clear that these numbers must be rational,
i.e., elements of the corresponding cyclotomic field
$\QQ(\xi_{n})$. This follows from the computability of the integrals
in equation \eqref{eq:auto} within these number fields.

The further restriction to algebraic \emph{integers}\/ is due to a
special structure of the frequency module of the tiling, i.e., the
$\ZZ$-span of the frequencies of all finite patches in the
tiling. Since our averaged quantities are simple integer linear
combinations of patch frequencies, a `quantisation' of the latter to
integers implies the result for the former. This phenomenon has been
observed before several times, and for various related problems
\cite{BGRJ,RRG,BG3,GB}. The proof relies on the topological structure
of the compact LI-class, viewed as a dynamical system under the
translation and/or inflation action \cite{AP,Bell,FHK}.

\subsection*{Acknowledgement}
It is our pleasure to thank Franz G\"{a}hler for valuable discussions.

\end{document}